\documentclass[sn-mathphys,Numbered]{sn-jnl}% Math and Physical Sciences Reference Style

\usepackage{graphicx}%
\usepackage{caption}
\usepackage{subcaption}
\usepackage{multirow}%
\usepackage{amsmath,amssymb,amsfonts}%
\usepackage{amsthm}%
\usepackage{mathrsfs}%
\usepackage[title]{appendix}%
\usepackage{xcolor}%
\usepackage{textcomp}%
\usepackage{manyfoot}%
\usepackage{booktabs}%
\usepackage{algorithm}%
\usepackage{algorithmicx}%
\usepackage{algpseudocode}%
\usepackage{listings}%
\usepackage{float}
\theoremstyle{thmstyleone}%
\theoremstyle{thmstyletwo}%

\theoremstyle{thmstylethree}%

\begin{document}

\title{Multi-Agent Stock Prediction Systems: Machine Learning Models, Simulations, and Real-Time Trading Strategies}

\author{\fnm{Daksh} \sur{Dave}}
\author{\fnm{Gauransh} \sur{Sawhney}}
\author{\fnm{Vikhyat} \sur{Chauhan}}

% \affil*[1]{Department of Electrical Electronics, BITS Pilani}
% \affil[2]{Department of Information System, College of Computers and Information Systems, Umm Al-Qura University, Al-lith-28434, Saudi Arabia}
% \affil[3]{}
% \affil[4]{Department of Computer Science, Quaid-i-Azam University, Islamabad, Pakistan}
% \affil[5]{Department of Computer Sciences, Technical University of Denmark, Denmark}

% \renewcommand\Authfont{\centering\large}
% \renewcommand\Affilfont{\centering\normalsize}

\abstract{This paper presents a comprehensive study on stock price prediction, leveraging advanced machine learning (ML) and deep learning (DL) techniques to improve financial forecasting accuracy. The research evaluates the performance of various recurrent neural network (RNN) architectures, including Long Short-Term Memory (LSTM) networks, Gated Recurrent Units (GRU), and attention-based models. These models are assessed for their ability to capture complex temporal dependencies inherent in stock market data. Our findings show that attention-based models outperform other architectures, achieving the highest accuracy by effectively capturing both short- and long-term dependencies. This study contributes valuable insights into AI-driven financial forecasting, offering practical guidance for developing more accurate and efficient trading systems.
}


\maketitle

\section{Introduction}
The stock market, characterized by its inherent volatility and complexity, poses a considerable challenge for accurate price forecasting and informed decision-making. In recent years, the rapid advancement of machine learning (ML) and deep learning (DL) techniques has brought transformative changes to the financial industry, offering more sophisticated tools for predicting market trends and developing trading strategies \cite{StockLiterature}. Traditional statistical methods, such as ARIMA \cite{StockArima} and linear regression \cite{StockLinear}, while foundational, have increasingly been repplaced by neural network-based models that excel in capturing complex temporal patterns in financial data. Among these, architectures like Long Short-Term Memory (LSTM), Gated Recurrent Units (GRU), and attention-based models have demonstrated superior performance in time-series prediction tasks, making them invaluable for stock price forecasting.

In this research, we systematically explore a broad array of ML and DL models, applying them to the task of stock price forecasting. Our study encompasses both univariate and multivariate time series data, providing a robust evaluation of several cutting-edge deep learning architectures. These include LSTM (and its variations such as Bidirectional LSTM and 2-Path LSTM), GRU (and its bidirectional and 2-path versions), as well as Seq2seq models and their extensions, such as Seq2seq VAE and hybrid approaches that integrate auto-encoders with boosting algorithms. Attention-based models, including the "Attention is All You Need" architecture, are also examined to assess their ability to enhance predictive accuracy.

Through case studies, particularly involving TESLA stock, we demonstrate the practical application of these models in real-world scenarios. The evaluation covers both predictive accuracy and their implications for automated trading strategies, providing a holistic view of the potential for AI-driven systems in the financial domain. The insights gained from this research contribute to the growing body of knowledge on AI-driven stock prediction systems and their capacity to revolutionize quantitative finance.

The major contributions of this paper are as follows: \begin{enumerate} \item A comprehensive evaluation of the effectiveness of various ML and DL models in forecasting stock prices. \item A practical demonstration of the models through case studies on real-world stock data, specifically focusing on TESLA. \item Insights into the potential of AI-driven systems to improve trading performance in stock markets. \end{enumerate}

The remainder of this paper is organized as follows: Section 2 presents the literature review, discussing prior work in the field. Section 3 describes the proposed methodology in detail. The experimental setup and datasets are outlined in Section 4. Finally, Section 5 discusses the results and their implications for future research.

\section{Literature Review}
Predicting the stock market is challenging yet crucial for investors, traders, and researchers. Various methods, including mathematical, statistical, and Artificial Intelligence (AI) techniques, have been proposed to forecast stock prices. Parmar et al. \cite{parmar2018stock} discovered that machine learning and deep learning prediction methods significantly outperform traditional stock market forecasting techniques in terms of both speed and accuracy. W. Khan \cite{khan2022stock} noted that social media significantly influences stock predictions, with Random Forest consistently performing well in all scenarios. Additionally, utilizing larger datasets and incorporating sentiment analysis into ML and DL models can enhance the precision of stock price predictions. These findings lead to development of many machine learning models, one being the Support Vector Machines (SVM) which belongs to the class of  supervised learning algorithms. Mankar et al. \cite{mankar2018stock} found that SVM proved to be a more effective and practical machine learning model for predicting stock price movements based on the sentiment expressed in tweets. Implementation of other deep learning techniques like artificial neural networks (ANNs) shows significant improvement over the previous machine learning solutions. Rao P. S. et al. \cite{rao2020survey} improved upon the existing model by proposing an artificial neural networks (ANNs), utilizing the backpropagation algorithm. The model outperformed the traditional regression methods and exhibited lower prediction errors. Naik N. and Mohan B. R. \cite{nikou2019stock} found that deep learning techniques outperformed machine learning techniques in terms of results. Further development into deep learning techniques suggested usage of RNN which can capture temporal dependencies and can further improve upon the prediction accuracy. 

Selvamuthu D., Kumar V., and Mishra A. \cite{selvamuthu2019indian} discovered that tick data provided more accurate predictions compared to 15-minute data. They reported that the Levenberg-Marquardt, Scaled Conjugate Gradient, and Bayesian Regularization methods achieved an impressive 99.9\% accuracy with tick data. Additionally, they suggested that using LSTM RNN would be advantageous. W. Fang et al. \cite{fang2019combine} proposed the LSTM predictive model, which includes an embedded layer and an automatic encoder. They note that stock news is not fully leveraged and that the approach has only been applied to the Chinese stock market. Their findings demonstrate that shallow machine learning algorithms, such as SVM and backpropagation, deliver lower accuracy compared to the LSTM and embedded LSTM methods. There have been several variations of LSTM RNN applications over stock prediction which proved it as a robust method over previous implementations. S. Mohan, S. Mullapudi, et al. \cite{mohan2019stock} demonstrated that RNN models incorporating LSTM outperformed the ARIMA model and showed a slight advantage over the Facebook Prophet algorithm. According to the research by Vignesh CK \cite{vignesh2018applying}, the LSTM method achieved a mean accuracy of 66.83 for Yahoo Finance, while the SVM method only attained an accuracy of 65.20. Training with smaller datasets while increasing the number of epochs can enhance testing outcomes across different datasets. M. Nikou, G. Mansourfar, and J. Bagherzadeh \cite{nikou2019stock} demonstrated that the LSTM-RNN block provides superior predictions of the closing price for the company's dataset compared to other methods. They recommend using combined models, such as the SVR model paired with a genetic algorithm, along with other hybrid approaches from machine learning algorithms. The findings of N. Sirimevan et al. \cite{sirimevan2019stock} indicates that the LSTM-RNN model, when combined with weighted average and differential evolution techniques, effectively predicted stock prices. 

A. Moghar and M. Hamiche \cite{moghar2020stock} demonstrated that the accuracy of stock price forecasting using LSTM models improves as the number of training epochs increases, specifically for GOOGL and NKE assets. J. Eapen, D. Bein, and A. Verma \cite{eapen2019novel} found that integrating Convolutional Neural Networks (CNN) with BiLSTM models enhanced stock market prediction accuracy by 9 percentage compared to using a single deep learning pipeline. Similarly, Adil Moghar and Mhamed Hamiche \cite{aguirre2018proceedings} concluded that BiLSTM models produce a lower Root Mean Squared Error (RMSE) compared to standard LSTM models, making them a more favorable option for stock prediction tasks. This hybrid approach also outperformed traditional Support Vector Machine (SVM) regressors in forecasting temporal sequences. Further optimizations have been proposed by incorporating external features into deep learning models to improve predictive accuracy. For instance, Khan, W., Malik, et al. \cite{khan2020predicting} discovered that adding sentiment analysis as an attribute had a minimal impact on stock price predictions but still improved the accuracy of machine learning algorithms by approximately 2 percentage. Moreover, M. Nabipour et al. \cite{nabipour2020predicting} found that using binary data instead of continuous data resulted in a significant enhancement in stock price predictions, highlighting the importance of data representation in predictive modeling. The reviewed literature highlights the growing effectiveness of deep learning models, such as LSTM and BiLSTM, in improving stock price forecasting accuracy through model integration and external optimizations. The reviewed literature highlights the growing effectiveness of deep learning models in improving stock price forecasting accuracy through model integration and external optimizations. However the field is evolving with time abd further research needs to be done to deploy machine learning models that can mimic humans and ensure that profit is generated with minimal human intervention.

% \cite{aguirre2018proceedings} demonstrated that employing Particle Swarm Optimization (PSO) enhances the accuracy of financial market predictions. 

% \begin{figure}[htbp]
%     \centering
%     \includegraphics[width = 0.9\textwidth]{Figures/METHODLOGYFRAMEWORL.drawio.pdf}
%     \caption{Proposed Framework For Anomaly Detection}
%     \label{fig:1}
% \end{figure}

\section{Proposed Methodology}
In this section, we outline the methodology for our experiment, which involves utilizing three distinct architectures: Long Short-Term Memory (LSTM), Gated Recurrent Units (GRU), and Transformers. The following sections will provide a detailed explanation of the implementation and processes for each architecture.

\subsection{LSTM}

LSTM (Long Short-Term Memory) \cite{lstm1997} is a type of Recurrent Neural Network (RNN) designed to capture long-term dependencies in sequential data and is suitable for processing and predicting with long intervals and delays in time series. LSTM can be used to make Stock Price Predictions where the network learns the temporal dependencies in historical stock price data, capturing different patterns to deduce future stock prices. The key feature of LSTM is its use of gating mechanisms to regulate the flow of information through the network, addressing the vanishing gradient problem that can occur in standard RNNs. The figure 1 denotes a LSTM cell.

\begin{figure}[htbp]
    \centering
    \includegraphics[width = 0.9\textwidth]{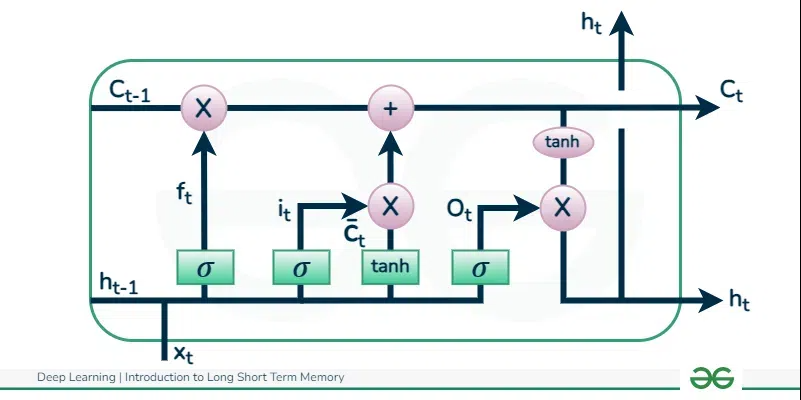}
    \caption{LSTM cell \cite{gfglstm}}
\end{figure}

An LSTM cell consists of three main gates:
\begin{enumerate}
    \item \textbf{Forget Gate}: Decides which part of the previous cell state should be forgotten.
    \item \textbf{Input Gate}: Determines which new information should be added to the cell state.
    \item \textbf{Output Gate}: Controls the output of the current cell.
\end{enumerate}
Additionally, an LSTM cell maintains a \textbf{cell state} \( C_t \), which acts as a memory, and a \textbf{hidden state} \( h_t \), which is used for the output at each time step.

At each time step \( t \), the LSTM performs the following operations:

1. \textbf{Forget Gate}: The forget gate determines how much of the previous cell state \( C_{t-1} \) should be retained for the current step:
\[
f_t = \sigma \left( W_f \cdot \left[ h_{t-1}, x_t \right] + b_f \right)
\]
Where:
- \( x_t \) is the input at time step \( t \),
- \( h_{t-1} \) is the hidden state from the previous time step,
- \( W_f \) and \( b_f \) are the forget gate weights and biases,
- \( \sigma \) is the sigmoid activation function, which outputs values between 0 and 1.

2. \textbf{Input Gate}: The input gate controls how much of the new information \( \tilde{C}_t \) should be added to the cell state:
\[
i_t = \sigma \left( W_i \cdot \left[ h_{t-1}, x_t \right] + b_i \right)
\]
Next, a candidate cell state \( \tilde{C}_t \) is computed:
\[
\tilde{C}_t = \tanh \left( W_C \cdot \left[ h_{t-1}, x_t \right] + b_C \right)
\]
Where:
- \( W_i \), \( b_i \), \( W_C \), and \( b_C \) are the input gate and candidate cell state weights and biases,
- \( \tanh \) is the hyperbolic tangent activation function.

3. \textbf{Cell State Update}: The new cell state \( C_t \) is a combination of the old cell state \( C_{t-1} \) (modulated by the forget gate) and the new candidate cell state \( \tilde{C}_t \) (modulated by the input gate):
\[
C_t = f_t \odot C_{t-1} + i_t \odot \tilde{C}_t
\]
Where \( \odot \) denotes element-wise multiplication.

4. \textbf{Output Gate}: The output gate determines what part of the cell state should be output as the hidden state \( h_t \):
\[
o_t = \sigma \left( W_o \cdot \left[ h_{t-1}, x_t \right] + b_o \right)
\]
Finally, the hidden state \( h_t \) is computed by applying the \( \tanh \) function to the cell state and modulating it with the output gate:
\[
h_t = o_t \odot \tanh(C_t)
\]

- The \textbf{forget gate} \( f_t \) determines how much of the previous memory \( C_{t-1} \) to retain.
- The \textbf{input gate} \( i_t \) modulates how much new information \( \tilde{C}_t \) to add.
- The \textbf{cell state} \( C_t \) is updated based on these two gates.
- The \textbf{output gate} \( o_t \) controls what is output as the hidden state \( h_t \).

By combining these gates, the LSTM cell can retain long-term dependencies over many time steps, which makes it effective for tasks involving sequential data, such as language modeling, time-series forecasting, and more.

\subsection{GRU}

GRU (Gated Recurrent Unit) \cite{gru2014} is a simpler variant of the LSTM (Long Short-Term Memory) network. It also addresses the vanishing gradient problem by using gating mechanisms but with a simplified structure. GRU can capture temporal dependencies in sequential data and is effective in predicting stock prices by learning from historical patterns and trends. A GRU has two gates: the \textbf{update gate} and the \textbf{reset gate}, which control the flow of information through the unit. The figure 2 denotes a GRU cell

\begin{figure}[htbp]
    \centering
    \includegraphics[width = 0.9\textwidth]{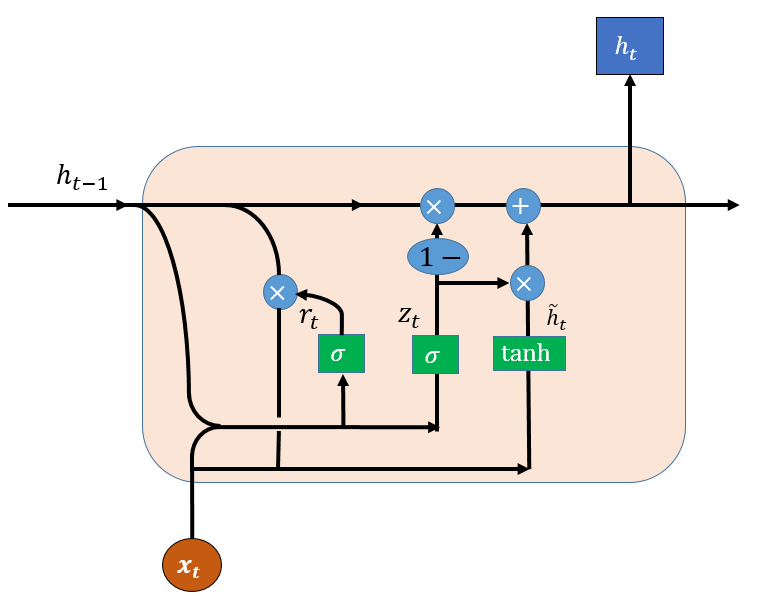}
    \caption{GRU cell \cite{huang2019convolutional}}
\end{figure}

At each time step \( t \), the GRU performs the following operations:

1. \textbf{Update Gate}: The update gate \( z_t \) determines how much of the previous hidden state \( h_{t-1} \) should be retained in the current hidden state:
\[
z_t = \sigma \left( W_z \cdot \left[ h_{t-1}, x_t \right] + b_z \right)
\]
Where:
- \( x_t \) is the input at time step \( t \),
- \( h_{t-1} \) is the hidden state from the previous time step,
- \( W_z \) and \( b_z \) are the weights and biases for the update gate,
- \( \sigma \) is the sigmoid activation function.

2. \textbf{Reset Gate}: The reset gate \( r_t \) controls how much of the previous hidden state \( h_{t-1} \) should be forgotten when updating the current hidden state:
\[
r_t = \sigma \left( W_r \cdot \left[ h_{t-1}, x_t \right] + b_r \right)
\]

3. \textbf{Candidate Hidden State}: The candidate hidden state \( \tilde{h}_t \) is calculated using the reset gate \( r_t \), which modulates the contribution of the previous hidden state:
\[
\tilde{h}_t = \tanh \left( W_h \cdot \left[ r_t \odot h_{t-1}, x_t \right] + b_h \right)
\]
Where:
- \( W_h \) and \( b_h \) are the weights and biases for the candidate hidden state,
- \( \odot \) represents element-wise multiplication.

4. \textbf{New Hidden State}: The new hidden state \( h_t \) is a combination of the previous hidden state \( h_{t-1} \) and the candidate hidden state \( \tilde{h}_t \), controlled by the update gate:
\[
h_t = z_t \odot h_{t-1} + (1 - z_t) \odot \tilde{h}_t
\]

- The \textbf{update gate} \( z_t \) controls how much of the past information is retained.
- The \textbf{reset gate} \( r_t \) determines how much of the previous hidden state should influence the current state.
- The \textbf{candidate hidden state} \( \tilde{h}_t \) is modulated by the reset gate.
- The final \textbf{hidden state} \( h_t \) is a blend of the previous hidden state and the candidate state, based on the update gate.

GRUs have fewer parameters than LSTMs because they do not have a separate memory cell, which often makes GRUs faster to train and perform comparably well in many tasks involving sequential data.

\subsection{Transformer}
\begin{figure}[htbp]
    \centering
    \includegraphics[height = 0.8\textwidth]{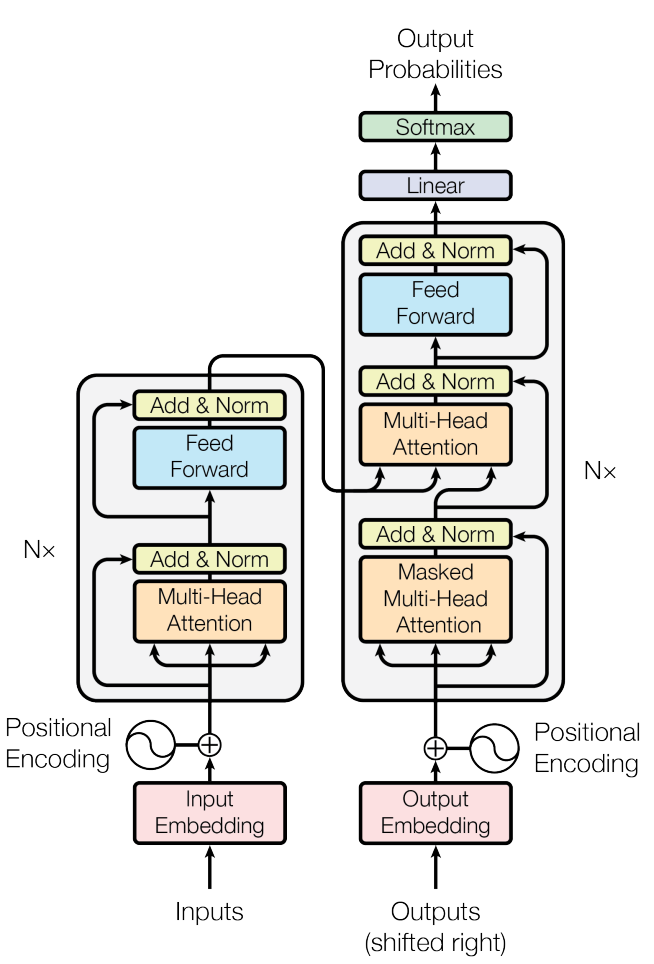}
    \caption{Transformer architecture}
    \label{fig:transformer architecture}
\end{figure}
The paper \textit{Attention is All You Need} \cite{vaswani2017attention} introduced the Transformer model, which relies solely on attention mechanisms to handle sequential data, without the need for recurrent or convolutional neural networks. The Transformer architecture has since become the foundation for many state-of-the-art models in natural language processing, such as BERT and GPT, and has also been applied to other domains, including time-series analysis and stock price prediction.

Unlike RNNs and GRUs, which process data sequentially, the Transformer processes the entire input sequence at once, allowing for better parallelization and faster training. This architecture is particularly useful for capturing long-range dependencies in the data. Figure~\ref{fig:transformer architecture} depicts the Transformer architecture.

The Transformer consists of an encoder and a decoder. For stock price prediction, you typically use only the encoder portion, which applies multi-head self-attention and feed-forward layers to learn the temporal patterns in the data. The attention mechanism allows the model to focus on different parts of the input sequence when making predictions. Since the Transformer lacks the inherent sequential structure of RNNs, positional encodings are added to the input data to provide information about the order of the time steps.

The attention mechanism in Transformers is defined by the following equation:

\[
\text{Attention}(Q, K, V) = \text{softmax}\left(\frac{QK^T}{\sqrt{d_k}}\right) V
\]

where:
\begin{itemize}
    \item \( Q \) (Query): The set of queries.
    \item \( K \) (Key): The set of keys.
    \item \( V \) (Value): The set of values.
    \item \( d_k \): The dimension of the key vectors, used to scale the dot product for numerical stability.
\end{itemize}

This equation computes a weighted sum of the values \( V \), where the weights are determined by the dot product between the queries \( Q \) and the keys \( K \), scaled by \( \sqrt{d_k} \), and passed through a softmax function to produce attention scores.

The attention mechanism in Transformers allows the model to capture both short- and long-term dependencies efficiently, making it highly suitable for tasks like stock price prediction where relationships between data points may span over long intervals.

\section{Experimental Design}

For our experiments, we plan to use the \textbf{Tesla Stock dataset}. A Tesla stock dataset typically contains historical data on the stock performance of Tesla Inc. (TSLA) over a given period of time. This dataset is often used for financial analysis, stock market prediction, machine learning models, and trend analysis. Here’s a general description of the common columns we found to be useful in the Tesla stock dataset:

\begin{itemize}
    \item \textbf{Date}: The date when the stock data was recorded.
    \item \textbf{Open}: The stock price at the opening of the trading day.
    \item \textbf{High}: The highest price reached during the trading day.
    \item \textbf{Low}: The lowest price reached during the trading day.
    \item \textbf{Close}: The stock price at the close of the trading day. This is often the most used column for trend analysis.
    \item \textbf{Adj Close}: The adjusted closing price, which accounts for corporate actions like stock splits, dividends, etc.
    \item \textbf{Volume}: The total number of shares traded during the day.
\end{itemize}

\begin{center}
\begin{table}[htbp]
    \begin{tabular}{|c|c|c|c|c|c|c|}
    \hline
    Date       & Open   & High   & Low    & Close  & Adj Close & Volume   \\ \hline
    2024-01-02 & 720.00 & 725.00 & 710.00 & 715.00 & 715.00    & 15,200,000 \\ \hline
    2024-01-03 & 715.50 & 730.00 & 705.50 & 725.00 & 725.00    & 17,500,000 \\ \hline
    \end{tabular}
    \caption{Sample rows from the Tesla Dataset}
    \label{tab:my_label}
\end{table}
\end{center}

The Table~\ref{tab:my_label} depicts sample rows from the Tesla dataset. For training and testing, we will partition the dataset into training and testing subsets. The training set comprises all data points from the initial timestamp up to 30 days prior to the final recorded timestamp. The test set includes the data from the last 30 days of the dataset. This approach ensures that the model is trained on historical data and is evaluated by forecasting based on the most recent 30-day period.

For each model in the experiments, the number of layers was set to 1. The size of the hidden layer was fixed at 128 units. Training was performed over 10 epochs and the learning rate was set to 0.01.

For the implementation, we use Python in a Jupyter notebook environment hosted on Google Colab and use L4 GPUs for training. The TensorFlow framework is employed to build and train the forecasting models. This setup provides an efficient, scalable environment for model training and experimentation, with access to GPU acceleration where necessary.

In addition to the baseline recurrent neural network models (LSTM and GRU), we explore multiple architectural variants designed to enhance predictive capabilities. One such variant is the \textbf{2Path} model, which processes the input sequence through two separate recurrent pathways before merging their outputs. This dual-branch design can capture multiple scales or perspectives of the time series. For instance, one path might focus on short-term volatility while the other attends to longer-term trends, and their combined embedding provides a richer context for the prediction task.

We also incorporate the \textbf{Seq2Seq} (Sequence-to-Sequence) framework, an encoder-decoder architecture that compresses an input sequence into a fixed-length latent representation and then generates the output sequence step-by-step from this representation. Originally developed for machine translation, Seq2Seq models are well-suited here for forecasting future stock prices, as they can handle variable-length input histories and produce multi-step forecasts that adapt to the complexity of financial time series.

To further enhance the model's ability to understand temporal dependencies, we leverage \textbf{Bidirectional} configurations. Bidirectional architectures process the input sequence in both forward and reverse directions, providing the model with future context when analyzing historical data. This approach mitigates the inherent limitation of unidirectional models, which only have access to past data points. By incorporating information from subsequent time steps, the model can develop a more nuanced and context-rich representation of the time series dynamics.

Lastly, we combine the advantages of both bidirectionality and the encoder-decoder paradigm in \textbf{Bidirectional-Seq2Seq} models. In this setup, the encoder is bidirectional, thereby extracting features informed by both past and future elements in the input sequence. The resulting latent representation is then utilized by the decoder, which generates forecasts step-by-step. The synergy of bidirectionality and Seq2Seq allows the model to fully leverage the temporal structure of the data, capturing intricate long-range dependencies and subtle market shifts that simpler architectures might overlook.

\section{Results and Discussion}

\subsection{Model Performance Overview}
The evaluation of various models is summarized in Table~\ref{tab:model_accuracy}. The table compares the accuracy of LSTM, GRU, their variants, and the Attention mechanism. The results highlight the differences in performance based on the model architecture.

\begin{table}[h!]
    \centering
    \begin{tabular}{l c l c}
        \toprule
        \textbf{Model Name 1} & \textbf{Accuracy 1} & \textbf{Model Name 2} & \textbf{Accuracy 2} \\
        \midrule
        LSTM                  & 89.2522             & LSTM-2Path            & 94.1620             \\
        LSTM-Seq2Seq          & 88.3902             & LSTM-Bidirectional    & 94.1620             \\
        LSTM-Bidirectional-Seq2Seq & 95.0921        & GRU                   & 84.1694             \\
        GRU-2Path             & 83.7746             & GRU-Seq2Seq           & 90.8855             \\
        GRU-Bidirectional     & 87.0331             & GRU-Bidirectional-Seq2Seq & 87.3672         \\
        \textbf{Attention}             & 95.1467             &                      &                     \\
        \bottomrule
    \end{tabular}
    \caption{Model Accuracy Comparison}
    \label{tab:model_accuracy}
\end{table}

Additional figures that support the results presented in this section are provided in the Appendix for further reference and analysis.
\subsection{Analysis of LSTM Variants}

\begin{figure}[h]
    \centering
    \includegraphics[width=\textwidth]{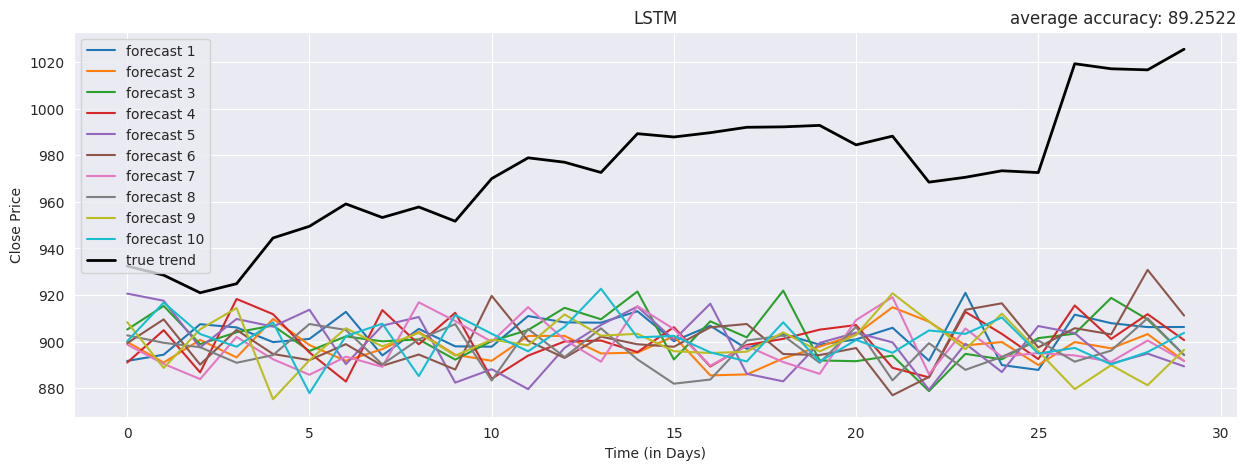}
    \caption{LSTM}
    \label{fig:lstm}
\end{figure}

\begin{figure}[h]
    \centering
    \includegraphics[width=\textwidth]{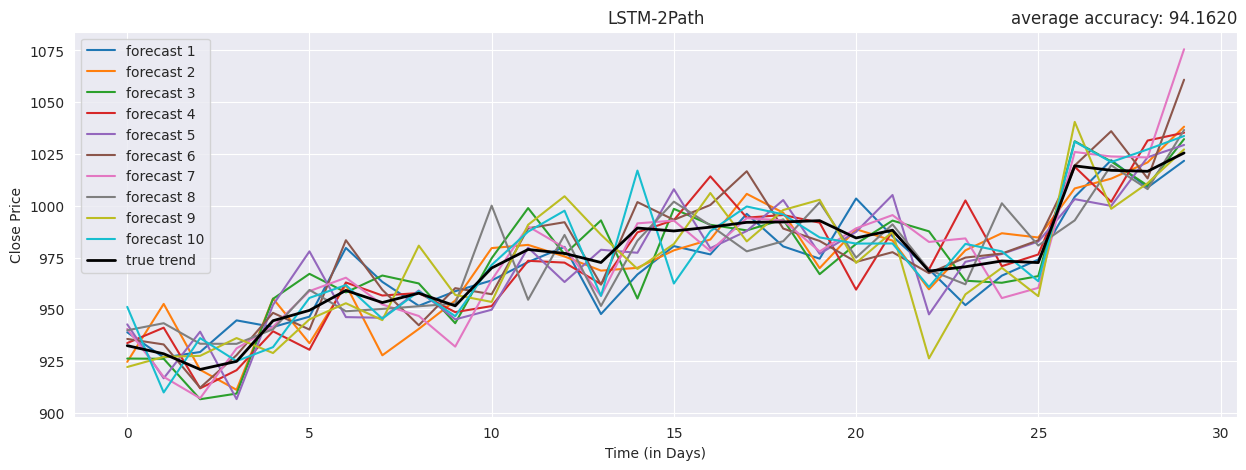}
    \caption{LSTM-2 Path}
    \label{fig:lstm_2path}
\end{figure}

\begin{figure}[h]
    \centering
    \includegraphics[width=0.9\textwidth]{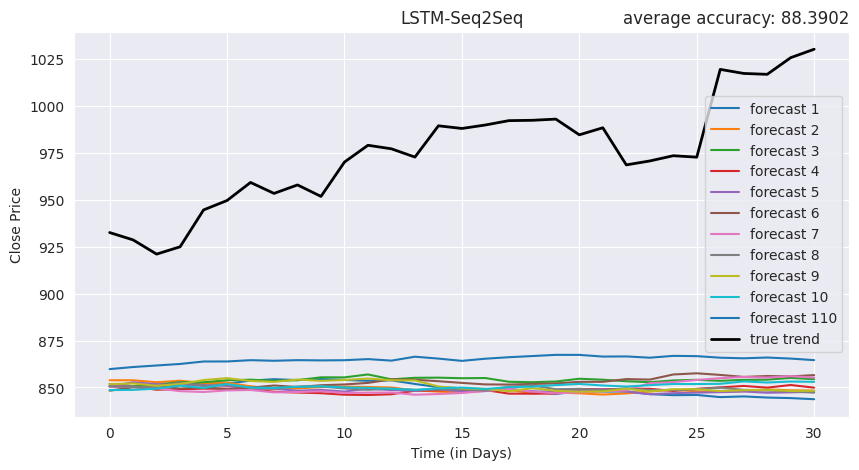}
    \caption{LSTM-Seq2Seq}
    \label{fig:lstm_seq2seq}
\end{figure}

\begin{figure}[h]
    \centering
    \includegraphics[width=\textwidth]{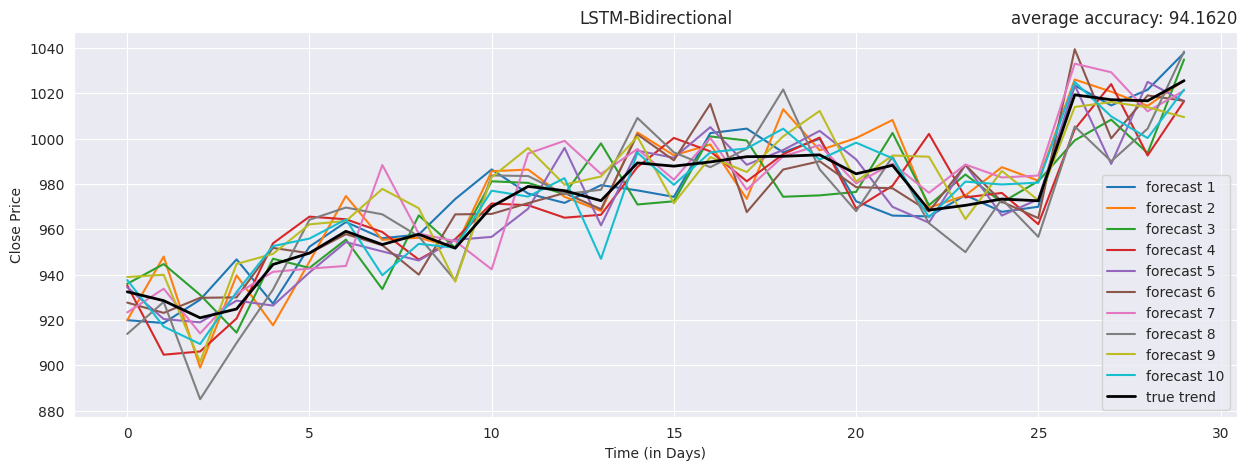}
    \caption{LSTM-Bidirectional}
    \label{fig:lstm_bidirectional}
\end{figure}

\begin{figure}[p]
    \centering
    \includegraphics[width=\textwidth]{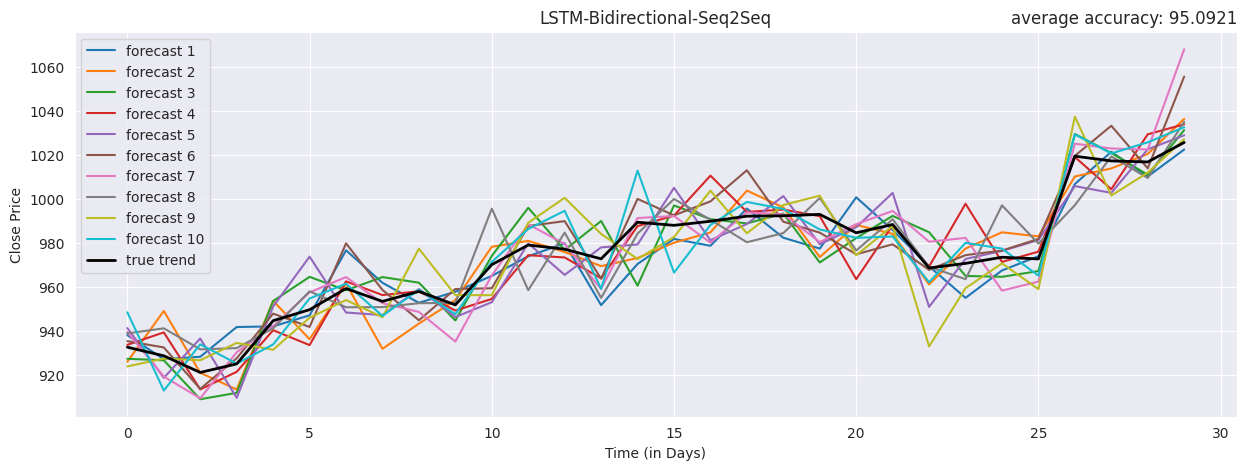}
    \caption{LSTM-Bidirectional-Seq2Seq}
    \label{fig:lstm_bidirectional_seq2seq}
\end{figure}

Figures~\ref{fig:lstm} to~\ref{fig:lstm_bidirectional_seq2seq} depict the performance trends of various LSTM-based models:

\begin{itemize}
    \item \textbf{LSTM} (Figure~\ref{fig:lstm}) achieved an accuracy of \textbf{89.2522\%}. Although it captures baseline temporal dependencies, it struggles with nuanced or abrupt changes in the input sequences.
    
    \item \textbf{LSTM-2Path} (Figure~\ref{fig:lstm_2path}) improved to \textbf{94.1620\%}. The inclusion of a two-path mechanism allows simultaneous processing of parallel data streams, enabling the model to capture more complex patterns that a single-path model might miss.
    
    \item \textbf{LSTM-Seq2Seq} (Figure~\ref{fig:lstm_seq2seq}) achieved \textbf{88.3902\%}. Contrary to expectations, integrating a seq2seq framework did not substantially enhance performance. One possible reason is that while seq2seq architectures shine in tasks like machine translation, this particular dataset may not benefit as strongly from the encoding-decoding paradigm.
    
    \item \textbf{LSTM-Bidirectional} (Figure~\ref{fig:lstm_bidirectional}) reached \textbf{94.1620\%}. This notable improvement underlines the importance of bidirectional processing, enabling the model to leverage information from both past and future temporal contexts to form a richer representation.
    
    \item \textbf{LSTM-Bidirectional-Seq2Seq} (Figure~\ref{fig:lstm_bidirectional_seq2seq}) outperformed all other LSTM models with \textbf{95.0921\%} accuracy. This configuration harnesses the strengths of both bidirectionality and a seq2seq architecture, indicating that when combined, these elements can more effectively model complex temporal dependencies.
\end{itemize}

\subsection{Analysis of GRU Variants}

\begin{figure}[p]
    \centering
    \includegraphics[width=\textwidth]{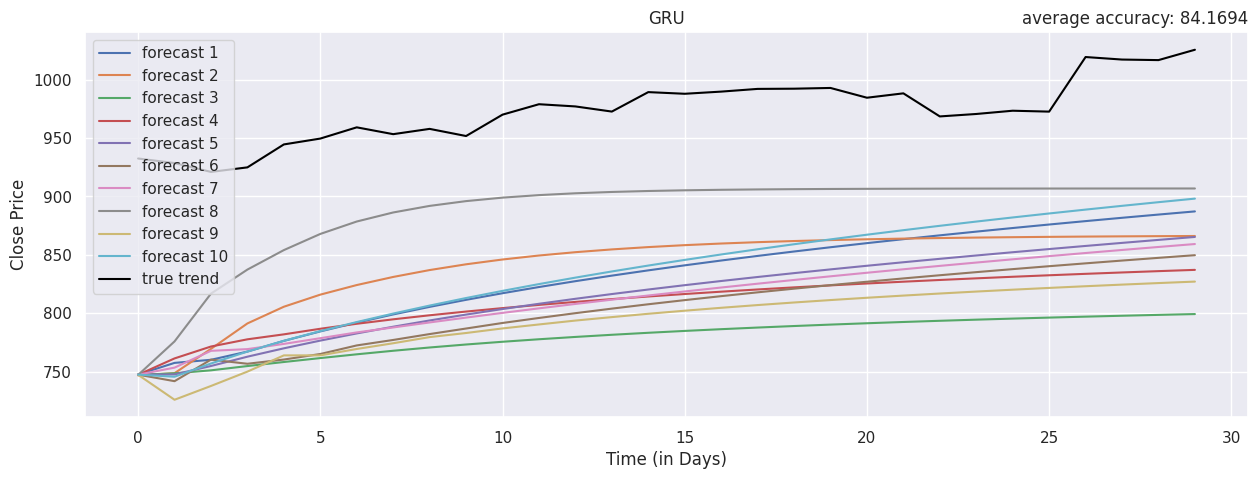}
    \caption{GRU}
    \label{fig:gru}
\end{figure}

\begin{figure}[p]
    \centering
    \includegraphics[width=\textwidth]{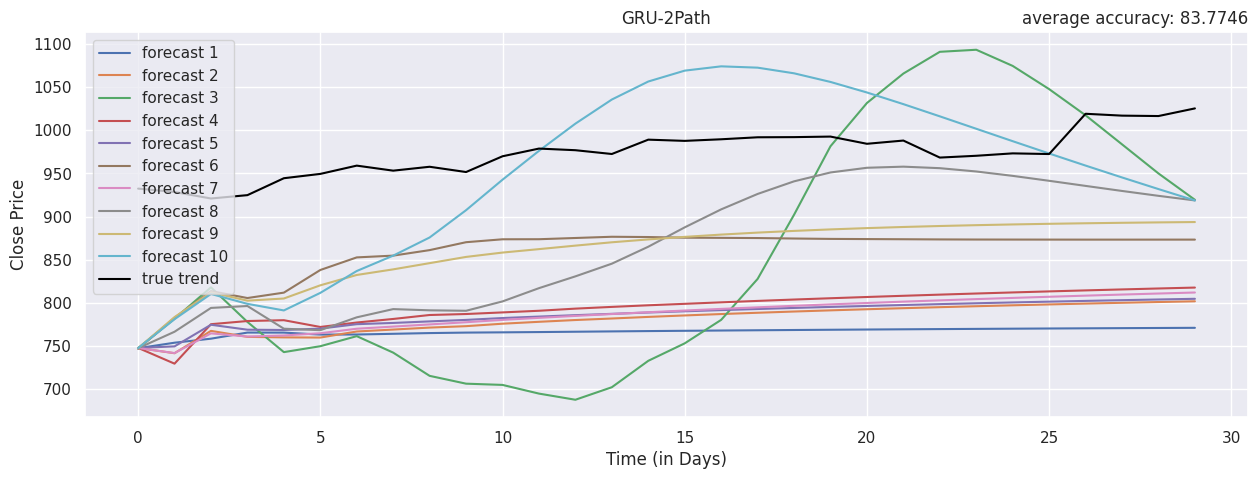}
    \caption{GRU-2Path}
    \label{fig:gru_2path}
\end{figure}

\begin{figure}[p]
    \centering
    \includegraphics[width=\textwidth]{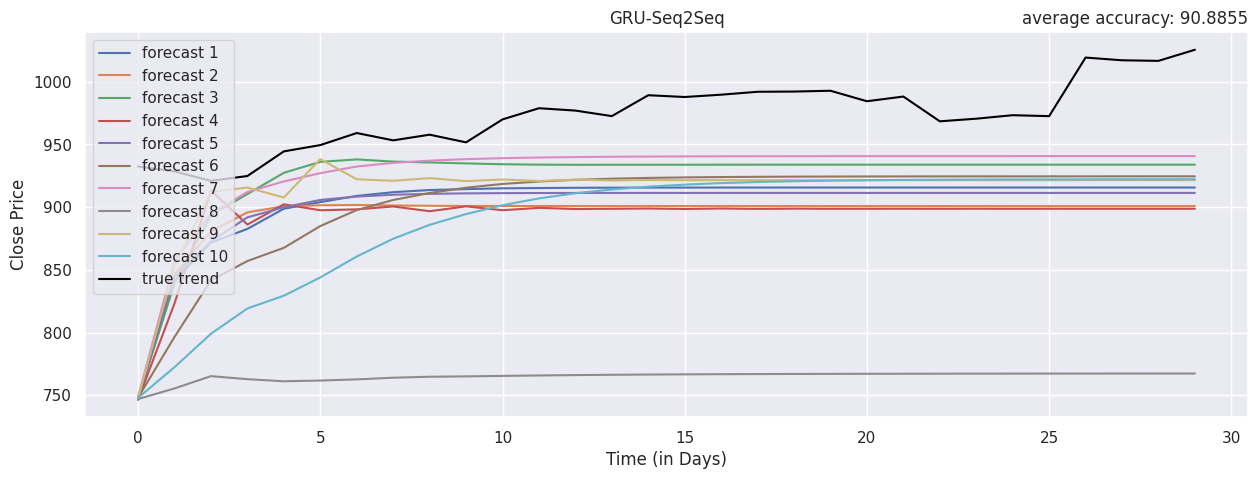}
    \caption{GRU-Seq2Seq}
    \label{fig:gru_seq2seq}
\end{figure}

\begin{figure}[p]
    \centering
    \includegraphics[width=\textwidth]{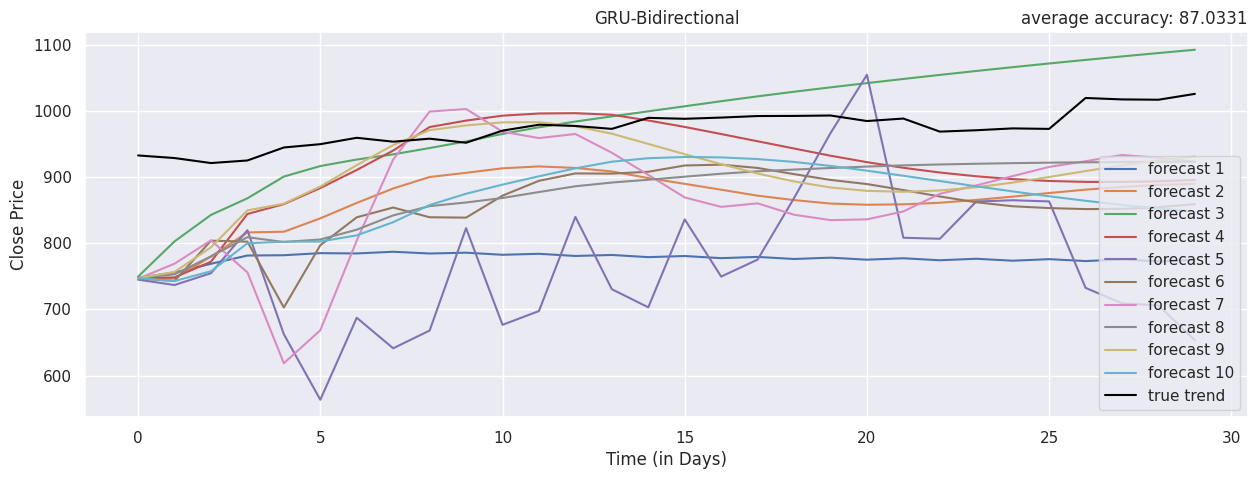}
    \caption{GRU-Bidirectional}
    \label{fig:gru_bidirectional}
\end{figure}

\begin{figure}[p]
    \centering
    \includegraphics[width=\textwidth]{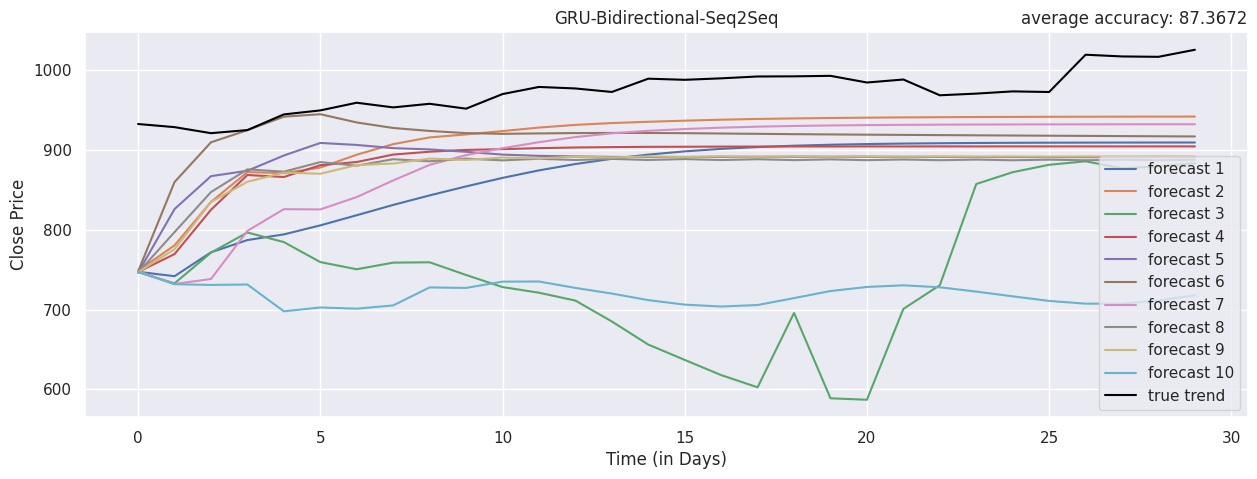}
    \caption{GRU-Bidirectional-Seq2Seq}
    \label{fig:gru_bidirectional_seq2seq}
\end{figure}

Figures~\ref{fig:gru} to~\ref{fig:gru_bidirectional_seq2seq} show that GRU-based models generally trail behind their LSTM counterparts. The simpler gating mechanism of GRUs, while computationally appealing, may not be as adept at handling more intricate long-term dependencies present in the data.

\begin{itemize}
    \item \textbf{GRU} (Figure~\ref{fig:gru}) obtained \textbf{84.1694\%} accuracy. While GRUs can capture basic temporal correlations, the baseline performance indicates limitations in handling more complex patterns.
    
    \item \textbf{GRU-2Path} (Figure~\ref{fig:gru_2path}) marginally decreased to \textbf{83.7746\%}, suggesting that the two-path strategy did not yield the same gains observed in the LSTM variant. The simpler GRU gating may not fully leverage parallel information streams as effectively as LSTM cells.
    
    \item \textbf{GRU-Seq2Seq} (Figure~\ref{fig:gru_seq2seq}) improved significantly to \textbf{90.8855\%}. The seq2seq structure appears beneficial here, likely because it provides a more flexible encoding-decoding dynamic that helps the GRU model better capture temporal dependencies.
    
    \item \textbf{GRU-Bidirectional} (Figure~\ref{fig:gru_bidirectional}) achieved \textbf{87.0331\%}, demonstrating that bidirectionality does provide a performance lift, albeit a more modest one than in LSTM models.
    
    \item \textbf{GRU-Bidirectional-Seq2Seq} (Figure~\ref{fig:gru_bidirectional_seq2seq}) yielded \textbf{87.3672\%}, a slight improvement over the bidirectional-only variant. While beneficial, the combination of bidirectionality and seq2seq for GRUs does not approach the accuracy gains seen with LSTMs, indicating that the intrinsic memory mechanisms of LSTMs may be better suited to exploiting these architectural enhancements.
\end{itemize}

\subsection{Attention Mechanism}

\begin{figure}[p]
    \centering
    \includegraphics[width=\textwidth]{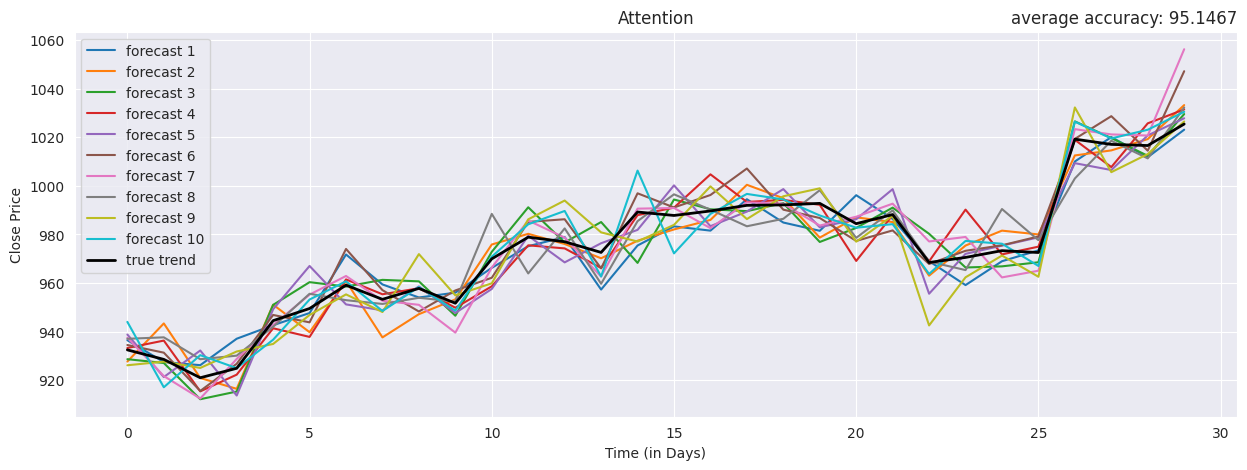}
    \caption{Attention Mechanism Visualization}
    \label{fig:attention}
\end{figure}

The \textbf{Attention} model (Figure~\ref{fig:attention}) attained the highest overall accuracy of \textbf{95.1467\%}. Unlike RNN-based models that rely on iterative hidden-state updates, attention mechanisms enable the model to directly “focus” on the most relevant portions of the input sequence. This ability to weigh and highlight critical time steps dynamically results in a stronger handling of long-range dependencies and subtle variations that may be difficult for standard recurrent architectures to capture adequately.

\subsection{Comparative Analysis and Reasoning}

\begin{itemize}
    \item \textbf{LSTM vs. GRU:} Across all configurations, LSTM-based models outperform their GRU counterparts. The richer gating mechanism of LSTMs seems better suited to extracting and preserving critical information over long sequences. As tasks become more complex, the LSTM’s capacity to handle intricate dependencies provides a tangible advantage.

    \item \textbf{Bidirectionality:} Introducing bidirectionality consistently enhances performance for both LSTM and GRU models. By processing input sequences in both forward and reverse temporal directions, the model obtains a more holistic view of the data. This expanded context is particularly valuable in tasks where future observations can inform the interpretation of earlier events.

    \item \textbf{Sequence-to-Sequence Architectures:} The seq2seq configurations yield mixed results. While LSTM-Bidirectional-Seq2Seq shows a notable improvement, the LSTM-Seq2Seq alone does not surpass the baseline. For GRUs, seq2seq integration markedly improves accuracy. These mixed outcomes suggest that seq2seq frameworks are beneficial primarily when combined with other architectural enhancements (e.g., bidirectionality) or when the base model can exploit the encoder-decoder paradigm effectively.

    \item \textbf{Attention:} The attention-based model emerges as the top performer. Its ability to directly attend to relevant parts of the input, rather than relying solely on hidden-state propagation, empowers it to capture complex, non-linear relationships. This proves especially valuable in datasets where essential cues are scattered over long time spans, and maintaining a single vector representation of the entire past may lead to information loss.
\end{itemize}

\section{Conclusion}
The results highlight the intricate nature of sequence modeling and the importance of selecting appropriate architectures to effectively capture temporal dynamics. While recurrent networks such as LSTMs and GRUs demonstrate a solid baseline for modeling sequential data, their performance can be significantly elevated through enhancements like bidirectionality, sequence-to-sequence (Seq2Seq) frameworks, and attention mechanisms.
Bidirectional processing enriches the model's contextual understanding by incorporating information from both past and future time steps. Seq2Seq architectures introduce flexibility by enabling the model to handle variable-length input and output sequences effectively. However, the most notable improvement arises from the incorporation of attention mechanisms, which allow the model to dynamically focus on the most relevant parts of the input sequence. This capability enables attention-based models to efficiently capture both short- and long-term dependencies, leading to state-of-the-art performance.
Overall, these findings underscore that augmenting recurrent networks with advanced architectural features is essential for decoding complex temporal patterns. The attention mechanism, in particular, stands out as a transformative element, offering a powerful means to overcome the limitations of traditional RNNs and unlocking superior predictive capabilities for tasks like stock price prediction.
Future work could explore hybrid models that integrate attention mechanisms with graph neural networks (GNNs) or transformers with domain-specific embeddings. Additionally, deploying these models in real-time trading environments and applying them to diverse datasets, including multi-asset portfolios, could further validate their effectiveness and generalizability.

\backmatter

\bibliography{sn-bibliography} % Common bib file
%% If required, the content of .bbl file can be included here once bbl is generated
%% \input sn-article.bbl
\newpage
\section{Appendix}

% Figure 2: LSTM-2 Path

% Figure 3: LSTM-Seq2Seq

% Figure 4: LSTM-Bidirectional

% Figure 5: LSTM-Bidirectional-Seq2Seq

% Figure 6: GRU

% Figure 7: GRU-2Path

% Figure 8: GRU-Seq2Seq

% Figure 9: GRU-Bidirectional

% Figure 10: GRU-Bidirectional-Seq2Seq

% Figure 11: Attention Mechanism

\end{document}